\newcommand{\PRE}[1]{}       % Use if journal style

\documentclass[aps,prl,twocolumn,amsmath,tightenlines,preprintnumbers,showpacs,amsfonts,amssymb]{revtex4}

\usepackage{bm}
\usepackage{epsfig}
\usepackage{graphicx}

\newcommand{\postscript}[2]{\setlength{\epsfxsize}{#2\hsize}
\centerline{\epsfbox{#1}}}
\newcommand{\lsim}{
\mathrel{\hbox{\rlap{\hbox{\lower4pt\hbox{$\sim$}}}\hbox{$<$}}}}
\newcommand{\gsim}{
\mathrel{\hbox{\rlap{\hbox{\lower4pt\hbox{$\sim$}}}\hbox{$>$}}}}

\begin{document}

\preprint{UFIFT-HEP-07-03}

\title{
\PRE{\vspace*{1.5in}} Revival of the Thermal Sneutrino Dark Matter
\PRE{\vspace*{0.3in}} }

\author{Hye-Sung Lee}
\author{Konstantin T.~Matchev}
\author{Salah Nasri}
\affiliation{Institute for Fundamental Theory, University of
Florida, Gainesville, FL 32611, USA } \PRE{\vspace*{.1in}}
\date{February 21, 2007}

\begin{abstract}
\PRE{\vspace*{.1in}} \noindent The left-handed sneutrino in the
Minimal Supersymmetric Standard Model (MSSM)
has been ruled out as a viable thermal dark matter candidate, due to
conflicting constraints from direct detection experiments and
from the measurement of the dark matter relic density. 
The intrinsic fine-tuning problem of the MSSM, however, motivates
an extension with a new $U(1)'$ gauge symmetry. 
We show that in the $U(1)'$-extended MSSM the {\em right-handed} 
sneutrino $\widetilde\nu_R$ becomes a good thermal dark matter candidate. 
We identify two generic parameter space regions where the 
combined constraints from relic density determinations, direct detection 
and collider searches are all satisfied.
\end{abstract}

%12.60.-i   Models beyond the standard model
%95.35.+d   Dark matter
%14.70.Pw   Other gauge bosons
\maketitle

Studies of the rotation curves of galaxies, large scale structures,
and recent measurements of the cosmic microwave background
radiation, have confirmed that about $23\%$ of the energy in the
Universe is in the form of cold dark matter (CDM) \cite{WMAP3}. The
origin and the nature of CDM is one of the biggest puzzles in both
particle physics and cosmology. Since all known particles are ruled 
out as dark matter candidates, dark matter provides the strongest 
phenomenological motivation for new physics beyond the Standard Model (SM).
The Minimal Supersymmetric Standard Model (MSSM),  
supplemented with an exact discrete symmetry ($R$-parity),
possesses two natural CDM candidates: the lightest neutralino
and the lightest scalar neutrino (sneutrino). The former is a 
generic mixture of the superpartners of the neutral gauge and Higgs 
bosons, and its phenomenology has been the subject of extensive 
studies over the last 20 years \cite{Jungman}. In contrast, the 
left-handed (LH) sneutrinos of the MSSM have been ruled out as a 
major component of the dark matter in the Universe, by the combination 
of cosmological and experimental constraints. More precisely,
LH sneutrinos are weakly charged, and typically annihilate too rapidly via
$Z$-mediated $s$-channel diagrams, resulting in a relic density too
small to account for all of the dark matter. To suppress the
annihilation rate it was proposed that the sneutrinos should be
either very light (${\cal O}({\rm GeV})$) \cite{lightSneutrino} or very
heavy (${\cal O}({\rm TeV})$) \cite{Olive}. However, a very light
sneutrino is excluded by the measurement of the invisible width of 
the $Z$ gauge boson, while a very heavy sneutrino is excluded by 
direct dark matter searches \cite{Olive}. Therefore, the LH sneutrinos 
of the MSSM are now disfavored as dark matter candidates.

On the other hand, the recent evidence of neutrino masses provides 
strong impetus for extending the particle content of the MSSM with 
right-handed (RH) neutrinos $\nu_R$ and their superpartners, the RH sneutrinos
$\widetilde\nu_R$. This opens up the new possibility that the dark matter
is due to a RH sneutrino, whose mass is plausibly in the TeV range. 
Indeed, if the neutrinos are Dirac, then a light RH neutrino  
is guaranteed, and its superpartner, whose mass is solely due to 
supersymmetry breaking effects, is expected to be around the Terascale.
Even if the neutrinos are Majorana particles, the smallness of their 
masses is naturally explained through a seesaw mechanism, thus
requiring the existence of RH Majorana neutrinos at some high scale, 
which could possibly be as low as the TeV scale. 
Whether or not the $\widetilde\nu_R$ is the lightest supersymmetric particle
(LSP) in the spectrum depends on the exact mechanism of supersymmetry 
breaking. In this letter we shall 
adopt a model-independent approach and simply assume that $\widetilde\nu_R$
is the LSP whose mass $M_{\widetilde\nu_R}$ is a free parameter. We shall 
then investigate the viability of $\widetilde\nu_R$ as a thermal dark matter 
candidate.

On the face of it, this idea cannot easily work, since the RH sneutrino 
is a SM singlet, and cannot be thermalized in the early universe through
SM gauge interactions.
One approach will be to assume that the $\widetilde\nu_R$'s
are produced non-thermally \cite{AD}, a scenario which is possible, 
but not very predictive. Another possibility is to take the LSP
as a mixture of LH and RH sneutrinos and adjust the
mixing angle to generate an acceptable thermal relic abundance
\cite{Arkani-Hamed}.
Here we shall pursue a different direction, namely, extending
the MSSM with an additional gauge symmetry, which would 
allow $\widetilde\nu_R$ to thermalize and then freeze-out with the 
proper relic abundance. For simplicity, we shall consider 
an extra Abelian gauge group $U(1)'$, under which all MSSM fields, 
as well as the RH sneutrino, are charged.
New Abelian gauge symmetries are predicted by many
new physics scenarios, including superstrings, 
extra dimensions, strong dynamics and grand unification.
The $U(1)'$-extended MSSM (UMSSM) \cite{Langacker} can also provide an elegant 
solution to the fine-tuning problem ($\mu$-problem \cite{muproblem}) 
of the MSSM when the symmetry is broken at TeV scale.
The UMSSM generically predicts a new gauge boson $Z'$ and
its superpartner, a $Z'$-ino $\widetilde Z'$, as well as 
a new singlet Higgs superfield $S$. All
of these new states are expected to have masses near the TeV scale.

Our setup is as follows. We assume three Dirac neutrinos, and 
correspondingly, three families of RH sneutrinos.
The allowed patterns of $U(1)'$ charges are singled out
by requiring that the $U(1)'$ be anomaly-free \cite{Anomaly}.
Even then, the model will have a large number of free parameters.
For simplicity, we shall make use of the $U(1)'$ charges
as predicted in $E_6$ grand unification. The $E_6$ group
contains two additional Abelian gauge groups,
$U(1)_{\chi}$ and $U(1)_{\psi}$.
Assuming only a linear combination of them at the TeV scale, 
the $U(1)'$ charge $Q'$ of any field is
given in terms of its $U(1)_{\chi}$ charge $Q_\chi$,
its $U(1)_{\psi}$ charge $Q_\psi$, and the 
mixing angle $\theta_{E6}$ as \cite{Langacker}
\begin{equation}
Q' = Q_\chi \cos\theta_{E6} + Q_\psi \sin\theta_{E6}\ .
\label{E6}
\end{equation}
This choice allows for tree-level neutrino Yukawa couplings
and neutrino mass generation through the usual Higgs mechanism.
Because of the smallness of the neutrino masses,
the L-R sneutrino mass mixing is extremely suppressed, and
the LH and the RH sneutrinos will be naturally decoupled.
We will assume that the LSP is the (almost) purely RH sneutrino in this letter, 
postponing the more general case of mixing with the LH sneutrino for a subsequent
publication.

In our numerical analysis, we further assume 
that any exotic chiral fields which might be required for anomaly
cancellation, are very heavy and will not affect the relic density
calculation. 
We shall take the value of the $U(1)'$ gauge coupling constant $g_{Z'}$
to be the GUT motivated value of $g_{Z'} = \sqrt{5/3} g_Y\equiv g_1$ where $g_Y$ is 
the gauge coupling constant of the hypercharge gauge group $U(1)_Y$.
We assume that the lightest RH sneutrino is sufficiently lighter than the other
two RH sneutrinos, and is the only dark matter candidate.
The generalization to the case of two or three degenerate RH sneutrino families, 
including the effects of coannihilations, is straightforward, using our 
results given below.

Due to the presence of the $U(1)'$ gauge interactions, in the early universe
the RH sneutrinos are in thermal equilibrium with the rest of the SM particles. 
As the temperature drops below $M_{\widetilde\nu_R}$,
they become non-relativistic and eventually freeze-out at some 
temperature $T_F$, following the usual scenario.
There are several relevant annihilation channels:
(1) $\widetilde Z'$-mediated $t$-channel processes 
$\widetilde \nu_R \widetilde \nu_R \to \nu \nu$, 
$\widetilde \nu_R^* \widetilde \nu_R^* \to \bar \nu \bar \nu$, and
$\widetilde \nu_R \widetilde \nu_R^* \to \nu \bar \nu$;
(2) $Z'$-mediated $s$-channel processes 
$\widetilde \nu_R \widetilde \nu_R^* \to f \bar f$ 
(in the final state we consider only the SM fermions,
including Dirac neutrino pairs);
(3) $\widetilde \nu_R$-mediated $t$-channel and 4-point diagram
$\widetilde \nu_R \widetilde \nu_R^* \to Z' Z'$, when $M_{Z'} <
M_{\widetilde \nu_R}$. We will not consider in this letter 
other possible channels such as annihilation into exotic fermions 
or Higgs bosons (through the $Z'$ resonance).

The present relic density of sneutrinos is found by solving the
Boltzmann equation and is given by
\begin{equation}
\Omega_{\widetilde\nu_R}h^2 \simeq \frac{1.04 \times 10^9\ {\rm GeV}^{-1}}{M_{Pl}}
\frac{x_F}{\sqrt{g_* (x_F)}} \frac{1}{a + 3 b / x_F} \label{omega}
\end{equation}
with
\begin{equation}
x_F \equiv \frac{M_{\widetilde \nu_R}}{T_F} = \ln \left( c
\sqrt{\frac{45}{8}} \frac{g_{\widetilde \nu_R}}{2 \pi^3}
\frac{M_{\widetilde \nu_R} M_{Pl} (a + 6 b/x_F)}{\sqrt{g_* (x_F)
x_F}} \right)\, , \label{freeze}
\end{equation}
where $M_{Pl} = 1.22\times 10^{19}~ {\rm GeV}$, $g_{\widetilde
\nu_R} = 1$, $c = 5/4$ and $g_*(x_F)$ is the total effective number
of relativistic degrees of freedom at freeze-out. In
Eqs.~(\ref{omega}) and (\ref{freeze}), we used the standard 
approximation $\left< \sigma v_{\rm rel} \right> = a + 6b/x_F$ 
for the thermally averaged annihilation cross-section times relative
velocity. Although this approximation is not very precise near thresholds 
and resonances, it provides a very good estimate of the cosmologically 
preferred values for the RH sneutrino masses. 
The leading contributions for each channel 
(either $a$-terms or $b$-terms) are given by:
\begin{widetext}
\begin{eqnarray}
a_{\nu \nu} &=& a_{\bar \nu \bar \nu} = g_{Z'}^4 Q'(\nu_R)^4 M_{\widetilde Z'}^2 / \left( \pi (M_{\widetilde Z'}^2 + M_{\widetilde \nu_R}^2 )^2 \right)\ , \\
b_{\nu \bar \nu} &=& g_{Z'}^4 M_{\widetilde \nu_R}^2 Q'(\nu_R)^2
\left( (M_{\widetilde Z'}^2 + M_{\widetilde \nu_R}^2 )^2
(Q'(\nu_L)^2 + Q'(\nu_R)^2) +
 2 (M_{\widetilde Z'}^2 + M_{\widetilde \nu_R}^2) (4 M_{\widetilde \nu_R}^2 - M_{\widetilde Z'}^2 ) Q'(\nu_L) Q'(\nu_R) \right. \nonumber \\
&& \left. + (-4 M_{\widetilde \nu_R}^2 + M_{Z'}^2)^2 Q'(\nu_R)^2 \right) / \left( 12 \pi (M_{\widetilde Z'}^2 + M_{\widetilde \nu_R}^2 )^2 \left| - 4 M_{\widetilde \nu_R}^2 + M_{Z'}^2 - i M_{Z'} \Gamma_{Z'} \right|^2 \right) \ ,\\
b_{f \bar f}&=& g_{Z'}^4 Q'(\nu_R)^2 (M_{\widetilde \nu_R}^2 - M_f^2)^{1/2} \left(4 M_{\widetilde \nu_R}^2 (Q'(f_L)^2 + Q'(f_R)^2) - M_f^2 (Q'(f_L)^2 - 6 Q'(f_L) Q'(f_R) + Q'(f_R)^2) \right) \nonumber \\
&&/ \left( 48 \pi M_{\widetilde \nu_R} \left| -4 M_{\widetilde \nu_R}^2 + M_{Z'}^2 - i M_{Z'} \Gamma_{Z'} \right|^2 \right)\ , \\
a_{Z'Z'} &=& g_{Z'}^4 Q'(\nu_R)^4 (M_{\widetilde \nu_R}^2 -
M_{Z'}^2)^{1/2} \left( 8 M_{\widetilde \nu_R}^4 - 8 M_{\widetilde
\nu_R}^2 M_{Z'}^2 + 3 M_{Z'}^4 \right) / \left( 16 \pi M_{\widetilde
\nu_R}^3 (- 2 M_{\widetilde \nu_R}^2 + M_{Z'}^2)^2 \right)\ ,
\end{eqnarray}
\end{widetext}
where $M_{Z'}$ ($\Gamma_{Z'}$) is the mass (width) of the $Z'$ gauge boson.

\begin{figure}[t!]
\postscript{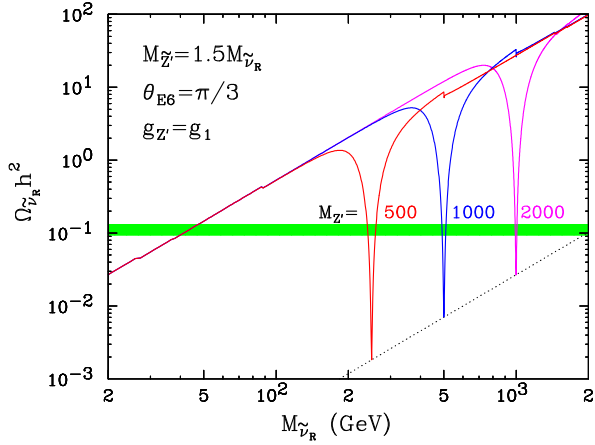}{0.9} \caption{Relic density 
$\Omega_{\widetilde\nu_R} h^2$ of the RH sneutrino versus its 
mass $M_{\widetilde\nu_R}$, for $\theta_{E6} = \pi/3$, $g_{Z'}=g_1$, 
and fixed $\widetilde Z'$ mass as $M_{\widetilde Z'} = 1.5 M_{\widetilde \nu_R}$.
Results are shown for three different values of $M_{Z'}$: 500 GeV (red),
1000 GeV (blue), and 2000 GeV (magenta). The shaded region
is the $2 \sigma$ range of $\Omega_{\rm CDM} h^2$ allowed by
WMAP+SDSS \cite{WMAP3}. 
The dotted line traces the minimum value of $\Omega_{\widetilde\nu_R}h^2$
on the $Z'$ resonance.
 \label{fig:relic} }
\end{figure}

Figure \ref{fig:relic} shows the relic density
$\Omega_{\widetilde\nu_R} h^2$ of the RH sneutrino
versus its mass $M_{\widetilde\nu_R}$, for  
$\theta_{E6} = \pi/3$, $g_{Z'}=g_1$, and
for fixed $M_{\widetilde Z'} = 1.5 M_{\widetilde \nu_R}$.
Results are shown for three different values of $M_{Z'}$: 500 GeV (red),
1000 GeV (blue), and 2000 GeV (magenta). The shaded region
is the $2 \sigma$ range of $\Omega_{\rm CDM} h^2$ allowed by
WMAP+SDSS $\Omega_{\rm CDM} h^2 = 0.111_{-0.015}^{+0.011}$
\cite{WMAP3}. The dotted line traces the minimum 
value of $\Omega_{\widetilde\nu_R}h^2$ on the $Z'$ resonance.
We see from Fig.~\ref{fig:relic} that over much of the
parameter space, the RH sneutrino relic density is too large 
and would overclose the Universe. This is expected, given the
absence of any SM interactions for the $\widetilde \nu_R$.
However, Fig.~\ref{fig:relic} also reveals the existence of 
at least two generic regions which yield acceptable values for
$\Omega_{\widetilde\nu_R} h^2$.
First, for the chosen values of the fixed parameters, 
there is a region around $M_{\widetilde \nu_R}=45$ GeV,
where $t$-channel annihilation through the relatively light
$\widetilde Z'$ is sufficient to reduce $\Omega_{\widetilde\nu_R} h^2$
to the desired values and below. In general, the location 
of this region (which is in a sense analogous to the ``bulk'' 
dark matter region of minimal supergravity) is given by
\begin{equation}
\frac{M_{\widetilde \nu_R}}{9\ {\rm TeV}}  \sim  g^2_{Z'}\, Q'(\nu_R)^2\,
\frac{r}{1+r^2}\, ,~~~~ r\equiv \frac{M_{\widetilde Z'}}{M_{\widetilde \nu_R}}\, .
\label{bulk}
\end{equation}
In addition, there is a $Z'$ resonance ``funnel'' region at 
\begin{equation}
M_{\widetilde \nu_R} \sim \frac{1}{2}\, M_{Z'}\ . 
\label{funnel}
\end{equation}
For the chosen 
values of the fixed parameters, this region is 
present over the whole range of sneutrino masses shown.
As the $Z'$ mass increases, however, the resonant dip
in $\Omega_{\widetilde\nu_R} h^2$ becomes more and more 
shallow, and eventually disappears for $M_{Z'}\gsim 4$ TeV
(with this choice of the fixed parameters).
Finally, a new channel $\widetilde\nu_R\widetilde\nu_R^*\to Z'Z'$ 
opens up for $M_{\widetilde \nu_R} > M_{Z'}$, 
as evidenced by the kinks at $M_{\widetilde \nu_R} \sim M_{Z'}$. 
With our choice of $E_6$ charge assignments, the $Z' Z'$
channel is unable by itself to satisfy the relic density
constraint, but may become relevant and provide a third 
good dark matter region if $g_{Z'}$ 
and/or the $U(1)'$ charges $Q'$ are assumed to be larger.

%%%%%%%%%%%%%%%%%%%%%%%%%%%% FIGURE 1 %%%%%%%%%%%%%%%%%%%%%%%%%%%%%%%%
%%%%%%%%%%%%%%%%%%%%%%%%%%%%%%%%%%%%%%%%%%%%%%%%%%%%%%%%%%%%%%%%%%%%%%%

The nucleus-dark matter interaction is given by the effective
Lagrangian,%&&{\cal {L}}_{\rm eff}(\widetilde\nu_R,q) =
\begin{eqnarray}
&&{\cal {L}}_{eff} =
i\frac{g^2_{Z'}}{M^2_{Z'}}Q'(\nu_R)\left(\widetilde
\nu_R^*\partial_{\mu}\widetilde \nu_R -
 \partial_{\mu}\widetilde\nu_R^*\widetilde\nu_R\right)\times\nonumber \\
 &&~ [\sum_{i=u,d}{Q'_V(q_i)\overline{q_i}\gamma_{\mu}q_i} +
 \sum_{i=u,d}{Q'_A(q_i)
 \overline{q_i}\gamma_{\mu}\gamma_5q_i}]
\end{eqnarray}
where $Q'_V(q_i)$ and $Q'_A(q_i)$ are the vector and axial charges
of the quark $q_i$, respectively. In the non-relativistic limit the
time component of the vector current dominates which gives the
spin-independent elastic scattering cross-section
\begin{equation}
\sigma_{\rm nucleon}^{\rm SI} = \frac{\lambda_N^2}{\pi A^2} \mu_n^2\ ,
\end{equation}
where $\mu_n$ is the effective mass of the nucleon and the
sneutrino, and $\lambda_N = Z\lambda_p + (A-Z)\lambda_n$, with
\begin{eqnarray}
\lambda_{p} &=& \frac{g_{Z'}^2}{M_{Z'}^2} Q'(\nu_R) \left[2 Q'_V(u)
+ Q'_V(d)\right],  \nonumber \\ \qquad \lambda_{n} &=&
\frac{g_{Z'}^2}{M_{Z'}^2} Q'(\nu_R) \left[2 Q'_V(d) +
Q'_V(u)\right].
\end{eqnarray}

%%%%%%%%%%%%%%%%%%%%%%%%%%%% FIGURE 2 %%%%%%%%%%%%%%%%%%%%%%%%%%%%%%%%
\begin{figure}[t!]
\postscript{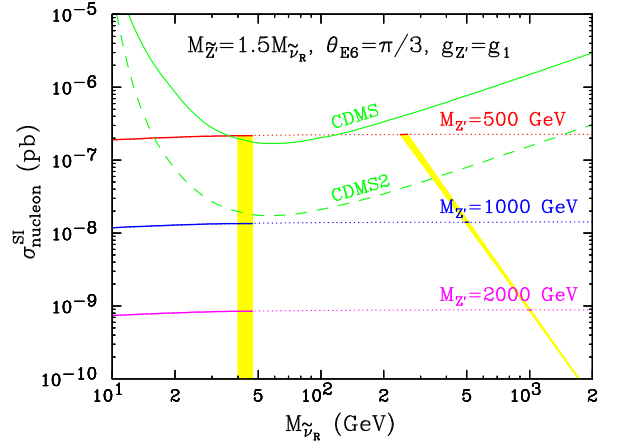}{0.9} \caption{Spin-independent
cross-section (normalized to a single nucleon) of the sneutrino-nucleus
interaction versus $M_{\widetilde\nu_R}$ for a Germanium type detector,
for the same parameter choices as in Fig.~\ref{fig:relic}.
The dotted (solid) portions of the curves correspond to 
unacceptable (acceptable) values for $\Omega_{\widetilde\nu_R} h^2$,
while in the yellow shaded region $\widetilde\nu_R$ can singlehandedly
explain all of the dark matter in the Universe. 
The green curves are the current (solid) and projected (dashed)
limits from the CDMS experiment \cite{CDMS2}. \label{fig:direct} }
\end{figure}
%%%%%%%%%%%%%%%%%%%%%%%%%%%%%%%%%%%%%%%%%%%%%%%%%%%%%%%%%%%%%%%%%%%%%%%

Figure \ref{fig:direct} shows our result for the spin-independent 
elastic scattering cross-section of the sneutrino dark matter 
in a Ge-type detector such as CDMS, for the same parameter 
choices as in Fig.~\ref{fig:relic}.
The solid (dashed) green curves are the current (projected for CDMS2) 
limits from the CDMS experiment \cite{CDMS2}.
The predicted cross-sections are almost flat over the whole range
$M_{\widetilde \nu_R} \gsim 10~ {\rm GeV}$ because $\mu_n \sim M_n=const$
for $M_{\widetilde \nu_R} \gg M_n$. The three curves 
are related by simple scaling, since $\sigma_{\rm nucleon}^{\rm SI}\sim M^{-4}_{Z'}$.
It is clear that by increasing the $Z'$ mass one can effectively 
suppress the elastic scattering cross-section and avoid
the direct detection constraint. 
However, even for the lowest $Z'$ masses allowed by the Tevatron, 
currently there is no direct detection constraint on $\widetilde\nu_R$ dark matter 
for the parameter choices in Figs.~\ref{fig:relic} and \ref{fig:direct}. 

%%%%%%%%%%%%%%%%%%%%%%%%%%%% FIGURE 3 %%%%%%%%%%%%%%%%%%%%%%%%%%%%%%%%
\begin{figure}[t!]
\postscript{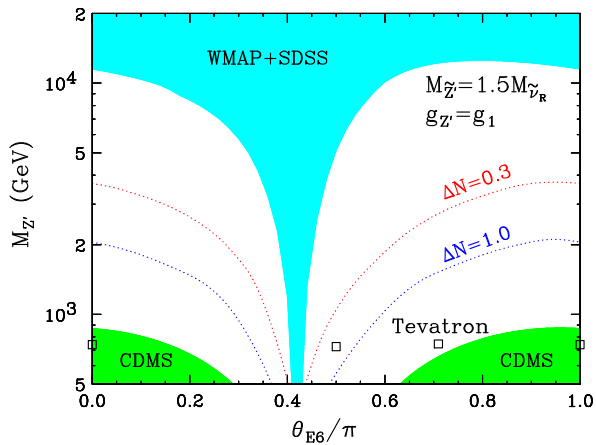}{0.9} \caption{
Experimental constraints on the ($\theta_{E6},M_{Z'}$) 
parameter space in the resonance funnel region 
$M_{\widetilde\nu_R}\sim M_{Z'}/2$, for fixed
$g_{Z'}=g_1$ and $M_{\widetilde Z'} = 1.5 M_{\widetilde \nu_R}$.
The upper (light blue) shaded region is cosmologically excluded,
while the lower (green) shaded region is currently ruled out by CDMS.
The squares indicate the most recent $Z'$ mass bounds from CDF 
\cite{Z2masslimit}. 
The dotted curves \cite{BBN} are the lower bounds on $M_{Z'}$ from the 
discrepancy in the $^4$He 
abundance, for an effective neutrino number of $\Delta N = 0.3$ (upper, red curve) 
and $\Delta N = 1$ (lower, blue curve), and for $T_c = 150~ {\rm
MeV}$.  The singular point $\theta_{E6} = 0.42 \pi$
corresponds to $Q'(\nu_R) = 0$. \label{fig:charge} }
\end{figure}
%%%%%%%%%%%%%%%%%%%%%%%%%%%%%%%%%%%%%%%%%%%%%%%%%%%%%%%%%%%%%%%%%%%%%%%

We have already seen that in the ``bulk'' $\widetilde\nu_R$ region
(\ref{bulk}) the relic density and the direct detection rates depend on
different parameters ($M_{\widetilde Z'}$ and $M_{Z'}$, respectively),
which to a large extent guarantees its viability.
Therefore, we shall now concentrate on the resonance ``funnel'' region
(\ref{funnel}), 
allowing also for variations in $\theta_{E6}$, and accumulate all relevant
experimental constraints in the ($\theta_{E6},M_{Z'}$) parameter plane. 
The results are shown in Fig.~\ref{fig:charge}, for fixed
$g_{Z'}=g_1$ and $M_{\widetilde Z'} = 1.5 M_{\widetilde \nu_R}$.
The upper (light blue) shaded region is cosmologically excluded:
we have already seen in Fig.~\ref{fig:relic}, that for any given value of 
$g_{Z'}$, $\theta_{E6}$ and $M_{\widetilde Z'}$, there is an upper limit 
on $M_{Z'}$, beyond which the relic density is too large, even on resonance.
The lower (green) shaded region is currently ruled out by the 
direct detection search at CDMS.
The dotted curves are the lower bounds on $M_{Z'}$ 
from requiring that the $^4$He abundance discrepancy be 
explained by light Dirac neutrinos coupled to $Z'$,
for two values of the effective neutrino number:
$\Delta N = 0.3$ (upper, red curve) 
and $\Delta N = 1$ (lower, blue curve), and for a choice of QCD
phase transition temperature $T_c = 150~ {\rm MeV}$ \cite{Steigman, BBN}.  
For a larger value of $T_c$, the curves will
be shifted to higher values of $M_{Z'}$. 
The Tevatron dilepton search provides typical bounds on the $Z'$ mass 
in the range $600 \sim 900~ {\rm GeV}$, depending on the $U(1)'$ 
charges \cite{Z2masslimit}. The squares indicate the most 
recent $Z'$ mass bounds from CDF within the $E_6$ model. 
The singular point $\theta_{E6} \sim 0.42 \pi$ corresponds to $Q'(\nu_R) \sim 0$,
when the RH sneutrino is (almost) decoupled and the Universe is overclosed.

The collider implications of the $\widetilde\nu_R$ LSP scenario 
are quite interesting, especially at hadron colliders such as the LHC. 
A sneutrino LSP would manifest itself as missing energy in the detector,
just like any other dark matter candidate. Since spin determinations at the
LHC are rather challenging, it is interesting to see whether this scenario 
can be discriminated from the usual case of neutralino LSP in supersymmetry, 
or its look-alike scenario of Universal Extra Dimensions \cite{UED}.
The scalar nature and/or new interactions of the dark matter particle 
also suggest interesting connections with other areas of cosmology, 
e.g. inflation \cite{U1_inflation}.

The prospects for indirect detection of $\widetilde\nu_R$ 
dark matter do not appear very promising, since the $a$ terms in most
of the annihilation channels are vanishing. 
One exception are the RH neutrino final states, which unfortunately 
lead to neutrino detection rates suppressed by the small Dirac 
neutrino mass. The other nonvanishing $a$-term is in the $Z'Z'$ final state, 
which only opens up for $M_{\widetilde \nu_R} > M_{Z'}$, and for the
typical values of the parameters considered here is rather small.

In this letter, we showed that in a natural extension of the MSSM 
with a new Abelian gauge symmetry $U(1)'$, the RH sneutrino $\widetilde\nu_R$
is a viable thermal dark matter candidate, satisfying all relevant experimental 
constraints. Our scenario is very generic and does not rely on the
particular choice of the $E_6$ charge assignments (\ref{E6}), or
the specific mechanism for solving the $\mu$-problem.
Our basic assumptions were just two: that there is a light RH neutrino 
whose superpartner gets its mass from supersymmetry breaking,
and that there are new gauge interactions at the TeV scale.
We then found two generic parameter space regions ((\ref{bulk}) and (\ref{funnel})) 
with good $\widetilde\nu_R$ dark matter. Considering the
effects of the additional Higgs singlets (either as particles 
in the final state or intermediate resonances), 
or the more general case of non-Abelian extra gauge symmetries,
will open up new and interesting possibilities for extending this scenario.

This work was supported by the Department of Energy under 
Grant No. DE-FG02-97ER41029.

%%%%%%%%%%%%%%%%%%%%%%%%%%%%%%%%%%%%%%%%%%%%%%%%%%%%%%

\end{document}